\documentclass[a4paper]{jpconf}
\usepackage{graphicx}
\usepackage{multirow}
\usepackage{hhline}
\usepackage{hyperref}
\usepackage[caption=false]{subfig}
\usepackage[export]{adjustbox}[2011/08/13]

\begin{document}

\title{Exploring compression techniques for ROOT IO}

\author{Z Zhang and B Bockelman}

\address{Holland Computer Center, University Nebraska $-$ Lincoln, Lincoln, NE 68588, USA}

\ead{zhan0915@huskers.unl.edu}

\begin{abstract}
ROOT provides an flexible format used throughout the HEP community. The number of use cases - from an archival data format to end-stage analysis - has required a number of tradeoffs to be exposed to the user. For example, a high ``compression level'' in the traditional DEFLATE algorithm will result in a smaller file (saving disk space) at the cost of slower decompression (costing CPU time when read). At the scale of the LHC experiment, poor design choices can result in terabytes of wasted space or wasted CPU time. We explore and attempt to quantify some of these tradeoffs. Specifically, we explore: the use of alternate compressing algorithms to optimize for read performance; an alternate method of compressing individual events to allow efficient random access; and a new approach to whole-file compression. Quantitative results are given, as well as guidance on how to make compression decisions for different use cases.
\end{abstract}

\section{Introduction}
\label{sec:introduction}

ROOT \cite{Brun199781} is an object-oriented C++ software framework developed at CERN.  It was designed to support the needs of particle physics data analysis and quickly grew to be the defacto storage format for HEP experiments.  One key is its flexibility: any instance of C++ objects can be efficiently stored into a ROOT file as machine-independent compressed format. The compressor reduces storage space and I/O bandwidth usage; it typically works column-wise, compressing similar objects together. However, it comes at a cost of increased CPU time when reading and writing files.  This tradeoff motivates us to give a quantitative comparison between different compression schemes for ROOT.

The rest of this paper is organized as followed: Section \ref{sec:background} gives background knowledge of ROOT. Section \ref{sec:algorithms} compares performance of three different compression libraries: ZLIB, LZMA and LZ4. Section \ref{sec:rac} discusses a new compression scheme - Random Access Compression (RAC) which provides very fast read performance in certain workload. Section \ref{sec:external} discusses External Compression which compresses ROOT files without any knowledge of data layout. 

\section{Background}
\label{sec:background}


The data generated in HEP are organized into events, each of which typically corresponds to a physical or simulated occurrence of particle collision.  Each event can be represented by multiple C++ objects in ROOT. ROOT usually uses a \texttt{TTree} object to represent an ordered collection of multiple events. \texttt{TTree} is further partitioned into branches called \texttt{TBranch}; each branch collects a set of similar objects from events (typically, objects of the same C++ type). When a C++ object is serialized, the resulting byte stream is stored in a memory buffer. Each branch contains one memory buffer.  When the buffer is full, ROOT compresses the data and a \texttt{TBasket} is created. \texttt{TBasket} is derived from \texttt{TKey} and contains additional information specific to the \texttt{TTree} navigation logic.  Hence, the serialized event data in a \texttt{TTree} consist of a sequence of \texttt{TBasket}s and their meta information.

ROOT uses a variant of ZLIB \cite{Deutsch:1996:ZCD:RFC1950} as its compressor. ZLIB uses dictionary compression algorithm called DEFLATE. Input data get compressed once a sliding window is filled. After compressing the current data, the window keeps sliding to successive blocks of input data and repeat the same process.  Each block is compressed using a combination of the LZ77 \cite{1055714} algorithm (searching for repeated substrings and replacing with back references) and Huffman coding \cite{4051119}.

\section{Comparison of different compression algorithms}
\label{sec:algorithms}

Besides ZLIB, ROOT also integrates LZMA \cite{lzma} as an alternative compressor. ZLIB provides fast compression and decompression speeds but it suffers from low compression ratio. LZMA, on the other hand, achieves high compression ratio but compression and decompression speeds are slow.

LZ4 \cite{lz4} is another compression library, which is reported as one order of magnitude faster than ZLIB, recently becomes attractive to the ROOT community. In this experiment, we compare the compression ratios, reading and writing speeds among these three compressors.



\subsection{Test Setup}
The file we tested is a CMS \cite{cms} file which contains 6500 events and 6.4 GB of data in total. Experiments are conducted on a VM with 4-core QEMU Virtual CPU version 1.5.3 and 8GB memory. The OS is Scientific Linux release 6.7. Both ZLIB and LZMA libraries can be configured with different compression levels ranging from 0 to 9. A higher level results in a better compression ratio. LZ4 library contains two sets of APIs: LZ4 and LZ4HC where HC stands for High Compression ratio. LZ4 does not need to set its level and LZ4HC is suggested to be configured between level 4 and 9. 

\subsection{Results}
As can be seen in Table \ref{tab:algrs}, LZMA gets the highest compression ratios. Comparing ZLIB, LZ4 improves the decompression speed by 6x to 8x. Let us take a close look at ZLIB-1 and LZ4HC-5, they have equivalent compression ratios. LZ4HC-5 takes more time (95.13s vs. 86.47s) to compress the file but only takes 13\% of decompression time of ZLIB-1. LZ4HC-9 achieves highest compression ratio among LZ4 families at 3.85. It is  less than ZLIB-5 (4.03) but only takes 2.54s to decompress the file versus 19.20s needed by ZLIB-5.


\begin{table}[!ht]
\centering
\caption{Comparison among compression algorithms}
\begin{tabular}{ccccc}
\hline
\multirow{3}{*}{Algorithm} & Compression & Decompression & Compressed  & Compression \\
 & Time (s) & Time (s) & File Size (GB) & Ratio \\
\hline
ROOT(ZLIB-6) & 228.67  & 18.45  & 1.54 & 4.16\\
ZLIB-1       & 86.47   & 21.51  & 1.79 & 3.58\\
ZLIB-5       & 159.84  & 19.20  & 1.58 & 4.05\\
ZLIB-9       & 1715.25 & 18.28  & 1.49 & 4.30\\
LZ4          & 11.26   & 2.97   & 2.17 & 2.95\\
LZ4HC-5      & 95.13   & 2.81   & 1.75 & 3.66\\
LZ4HC-9      & 275.31  & 2.54   & 1.66 & 3.86\\
LZMA-1       & 823.84  & 230.64 & 1.35 & 4.74\\
LZMA-5       & 3318.35 & 211.71 & 1.23 & 5.20\\
LZMA-9       & 4969.20 & 212.47 & 1.21 & 5.29\\
\hline
\end{tabular}
\label{tab:algrs}
\end{table}

\newcommand{\tabracratio}{
\begin{tabular}{ccccccc}
\hline
\multirow{2}{*}{} & \multirow{2}{*}{Raw File(GB)} & \multicolumn{2}{c}{Compressed File(GB)} & \multicolumn{2}{c}{Compression Ratio} & \multirow{2}{*}{$\frac{w/ RAC}{w/o RAC}$} \\
\hhline{~~----~}
 & & w/o RAC & w/ RAC & w/o RAC & w/ RAC & \\
\hline
All Branches  & 11.49 & 2.12 & 3.67 & 5.43 & 3.13 & 1.73\\
TFloat Branch &  4.01 & 0.79 & 2.27 & 5.09 & 1.77 & 2.88\\
TSmall Branch &  3.75 & 0.67 & 0.74 & 5.58 & 5.05 & 1.10\\
TLarge Branch &  3.73 & 0.66 & 0.66 & 5.69 & 5.69 & 1.00\\
\hline
\end{tabular}
}

\newcommand{\tabractime}{
\begin{tabular}{ccc}
\hline
               & w/o RAC & w/ RAC \\
\hline
Real Time (s) & 273.08 & 670.32 \\
CPU Time (s)  & 240.68 & 657.69 \\
\hline
\end{tabular}
}

\newcommand{\tabracreadrand}{
\begin{tabular}{ccccccccc}
\hline
\multirow{2}{*}{} & \multicolumn{4}{c}{w/o RAC} & \multicolumn{4}{c}{w/ RAC} \\
\hhline{~--------}
 & \multicolumn{2}{c}{Cold Cache} & \multicolumn{2}{c}{Hot Cache} & \multicolumn{2}{c}{Cold Cache} & \multicolumn{2}{c}{Hot Cache} \\
\hhline{~--------}
 & RT(s) & CT(s) & RT(s) & CT(s) & RT(s) & CT(s) & RT(s) & CT(s) \\
\hhline{---------}
TFloat   & 9.498 & 4.553 & 3.985 & 3.980 & 21.011 & 2.563 & 0.843 & 0.840 \\
TSmall & 9.004 & 2.533 & 1.367 & 1.363 & 11.972 & 0.927 & 0.284 & 0.287 \\
TLarge & 2.308 & 1.153 & 0.967 & 0.970 & 2.123  & 1.117 & 1.005 & 1.007 \\
\hline
\end{tabular}
}

\newcommand{\tabracreadseq}{
\begin{tabular}{ccccccccc}
\hline
\multirow{2}{*}{} & \multicolumn{4}{c}{w/o RAC} & \multicolumn{4}{c}{w/ RAC} \\
\hhline{~--------}
 & \multicolumn{2}{c}{Cold Cache} & \multicolumn{2}{c}{Hot Cache} & \multicolumn{2}{c}{Cold Cache} & \multicolumn{2}{c}{Hot Cache} \\
\hhline{~--------}
 & RT(s) & CT(s) & RT(s) & CT(s) & RT(s) & CT(s) & RT(s) & CT(s) \\
\hhline{---------}
TFloat   & 97.370 & 95.193 & 94.507 & 94.507 & 92.734 & 90.570 & 91.664 & 91.660 \\
TSmall & 12.070 & 11.337 & 10.915 & 10.907 & 11.989 & 4.880  & 3.584  & 3.577  \\
TLarge & 13.376 & 10.347 & 9.444  & 9.440  & 21.004 & 10.926 & 9.523  & 9.520  \\
\hline
\end{tabular}
}

\section{Random Access Compression (RAC)}
\label{sec:rac}

Some physicists wish to compress data in fine-grained units \cite{188994} so that users can improve read performance by only decompressing partial data. By default, ROOT requires all objects in a TBasket buffer to be compressed at once; we introduce a ``RAC" mode, compressing each event within the buffer individually. There are two downsides of RAC: poor compression ratio and some overhead added to TBasket. Since an event in RAC is independent with prior events, RAC can not search strings in previous events for a match and thus loses compression ratio. Additionally, in order to keep track of access points of different events in a compressed basket, we add an array in TBasket and store event offsets. When reading a particular event, this technique allows us to find the access point where the event is located within the TBasket buffer and start decompressing it from there. 

\subsection{Test Setup}

We create some dummy events to simulate typical cases. We create three types of events: TFloat, TSmall and TLarge. TFloat is a tiny event consisting of 6 Floating Points(FPs) with the same value. A TSmall contains 1000 FPs. To construct a TSmall, we first randomly generate a FP and then repeat this value 6 times. We start over the same process until all 1,000 FPs are filled. A TLarge is generated in the same way but contains 1,000,000 FPs. Each event of TFloat, TSmall and TLarge is 39 bytes, 4KB and 4MB respectively.

In the experiment, we create a ROOT tree containing three branches, each of which stores TFloats, TSmalls and TLarges respectively. We iterate a loop 1,000 times. During each iteration, one TLarge, 1,000 TSmalls and 100,000,000 TFloats are created and filled into their corresponding TBranches. The reason of choosing these numbers is to make sure each branch contains approximately the same amount of data.

\subsection{Results}

We write our generated events to a file and then read each individual event back from the file. We measure the performance from three aspects: compression ratio, read time and write time. When we measure read time, we test two different workloads: random reads and sequential reads. 

\begin{figure}[!ht]
\centering
\captionsetup[subfloat]{position=top}
\subfloat[Comparison of compression ratios]{\tabracratio} \\
\subfloat[Time spent on compression]{\tabractime} \\[-0.1ex]
\captionsetup[subfloat]{position=bottom}
\subfloat[Figure representation of compression ratios]{\includegraphics[height=3in, width=5in]{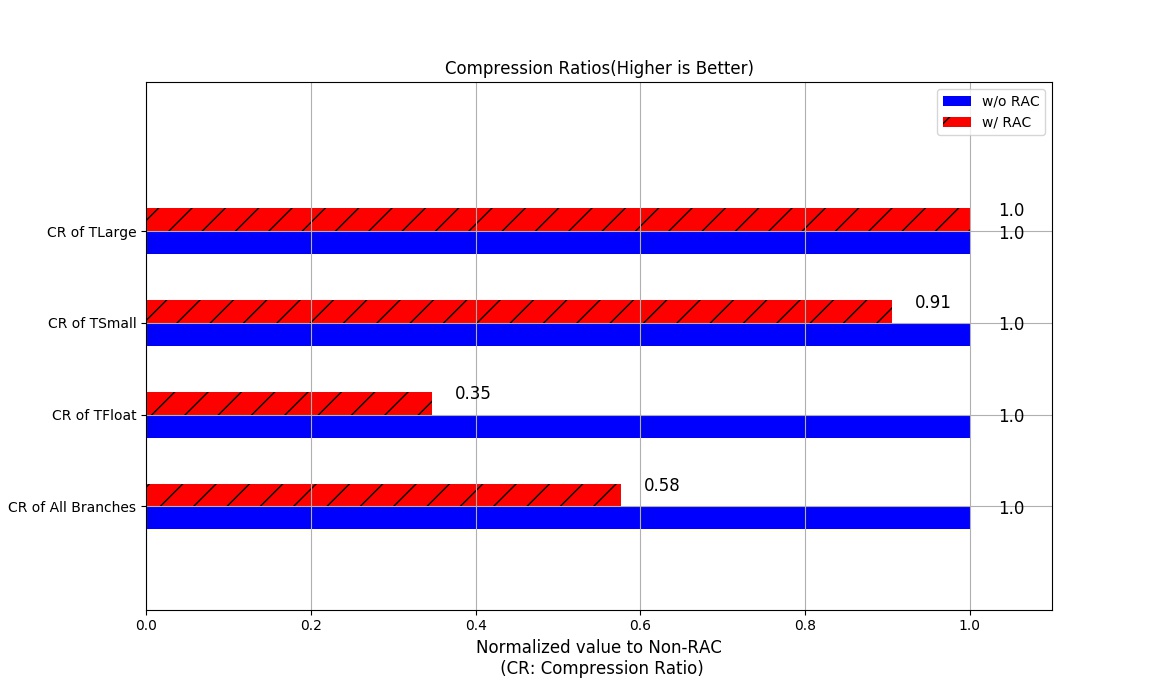}\label{fig:racratio}}
\caption{Compression ratios and write time of RAC on different events}
\label{mix:racratio}
\end{figure}

Figure \ref{mix:racratio} shows the compression ratios of different types of events. With original ZLIB TFloat can achieve a compression ratio of 5.09 while with RAC it only has a compression ratio of 1.77. The intuition behind is that the original design of the compressor accumulates 64KB of TFloats in a basket and compresses the entire basket at once. In contrast, RAC only compresses each TFloat independently and thus it can not discover the redundancy between different TFloat objects. In addition, RAC needs to store an access offset for each TFloat. This incurs a significant amount of overhead when dealing with small events and further downgrades the compression ratio. Regarding to TSmall, the compression ratio gets slightly smaller when we apply RAC. Each TSmall is 4 KB and there should be approximately 16 events in a 64 KB basket. Although there are more data to look at within a buffer, based on how we generate an event, two FP numbers with their distance longer than 6 should look like random. It does not make difference anymore once we compress a group of events larger than 6. For the case of TLarge, since the size of each event already goes beyond the basket size, therefore each basket can only hold a single event and RAC does not make any difference in this case.

\begin{figure}[!ht]
\centering
\captionsetup[subfloat]{position=top}
\subfloat[Time spent on decompression]{\tabracreadrand} \\[-0.1ex]
\captionsetup[subfloat]{position=bottom}
\subfloat[Figure representation of the above table]{\includegraphics[height=4.5in, width=7in, center]{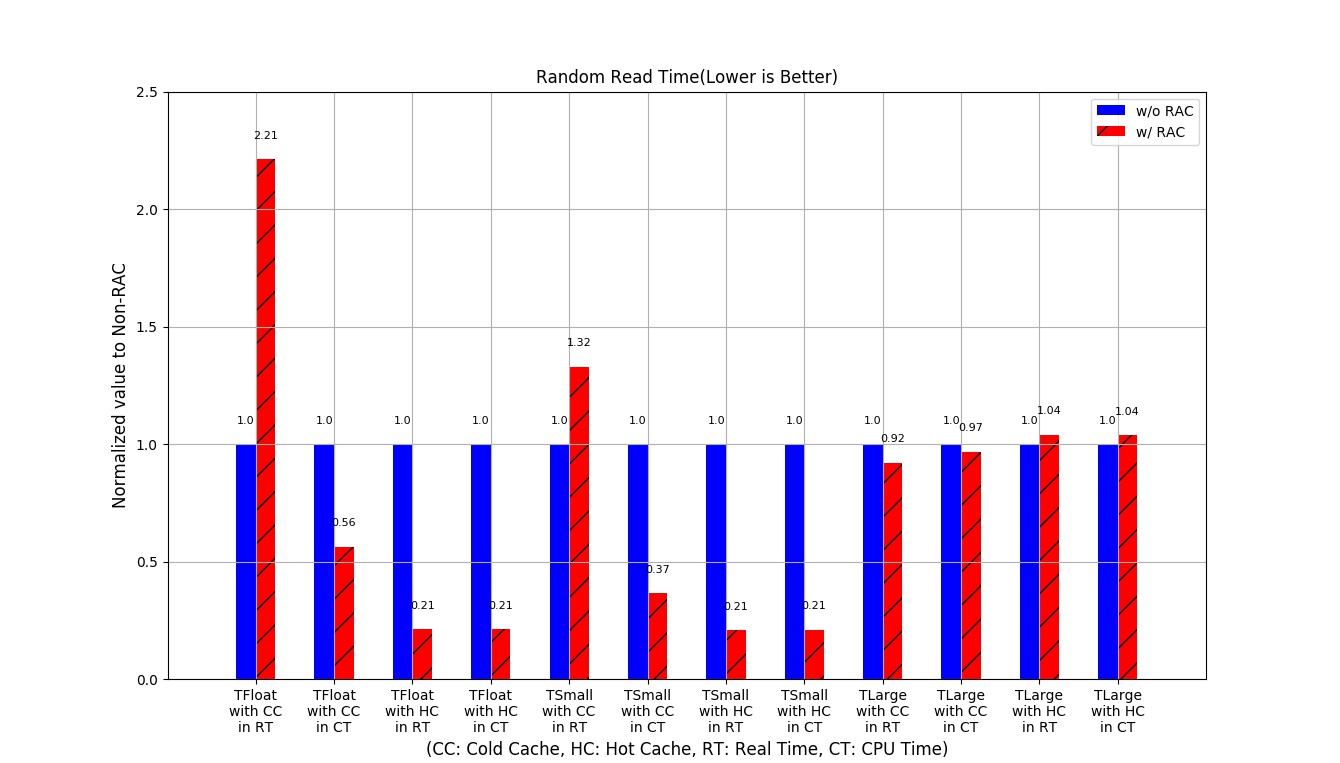}\label{fig:racreadrand}}
\caption{RAC read time of random reads}
\label{mix:racreadrand}
\end{figure}

We also evaluate random read performance. We run two types of tests. The first test is conducted by randomly reading events after completely cleaning the cache. We call it ``cold cache''. The second test randomly reads a file which is preloaded into memory. we call it ``hot cache''. The result is shown in Figure \ref{fig:racreadrand}. Intuitively, time spent on reading small events is dramatically reduced with RAC because it only needs to decompress very small portion of the basket. Instead, ZLIB has to decompress the whole basket before reading the event. We randomly read 1000 events from each of three event branches. As we expect, randomly reading TFloats and TSmalls with RAC only takes 21\% of the CPU time (with hot cache) comparing to ZLIB. Surprisingly, RAC takes more time to read TLarge events than ZLIB. This is due to the design overhead of adding more data structures to TBasket.

\begin{figure}[!ht]
\centering
\captionsetup[subfloat]{position=top}
\subfloat[Time spent on decompression]{\tabracreadseq} \\[-0.1ex]
\captionsetup[subfloat]{position=bottom}
\subfloat[Figure representation of the above table]{\includegraphics[height=4.5in, width=7in, center]{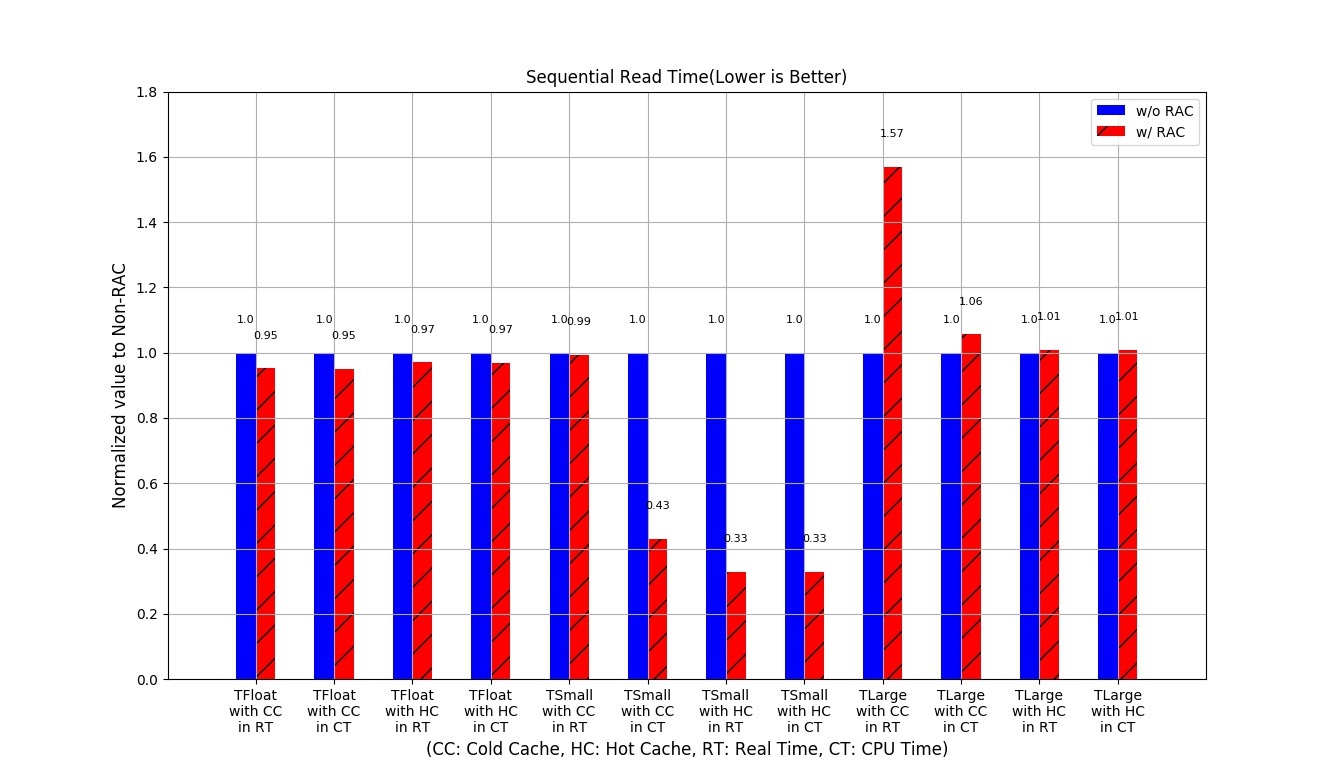}\label{fig:racreadseq}}
\caption{RAC read time of sequential reads}
\label{mix:racreadseq}
\end{figure}

We also test sequential reads. The result is shown in the Figure \ref{fig:racreadseq}. TSmall events have significant improvement if we use RAC. It is because the RAC saves time on decompression, therefore the bottleneck of the program starts shifting from CPU-bounded to IO-bounded. Based on our observation, in the cold cache case, it takes 12.07 (real time) seconds to sequentially read all TSmall events in the file without RAC and 11.99 (real time) seconds for the file with RAC.  However, it only takes 4.88 seconds (CPU time) to decompress the data with RAC. It still takes 11.34 seconds (CPU time) to decompress data without RAC. This implies, with RAC, the whole read process is still bounded by IO. The IO bound disappears when we read from hot cache, we can see the reading speed of TSmall events are much faster if we use RAC.

\section{External Compression}
\label{sec:external}

We compare the performance of ``external compression" - compressing the file with a separate program outside ROOT.  Here, we blindly divide ROOT files into smaller blocks of the same size and compress each individually. We use SquashFS \cite{squashfs} to test this approach.  Importantly, it presents the file as a mounted filesystem - entries in the page cache are \textit{uncompressed} buffers.

There are two purposes of doing external compression. First, we want to give a quantitative comparison of how well ROOT formats the raw data. Second, we want to verify if user-space compression is efficient.

\subsection{Test Setup}

In this experiment, we use the same ROOT file as in Section \ref{sec:algorithms}. We mount uncompressed file as SquashFS. SquashFS supports ZLIB and the default compression level is 9. We use the same compression settings in ROOT. We tried different block sizes for SquashFS. A block of SquashFS is an interchangeable term for a basket in ROOT. We also examine various basket sizes in ROOT corresponding to SquashFS's block sizes. 

\begin{figure}[!ht]
\begin{center}
\includegraphics[height=2.5in, width=3.5in]{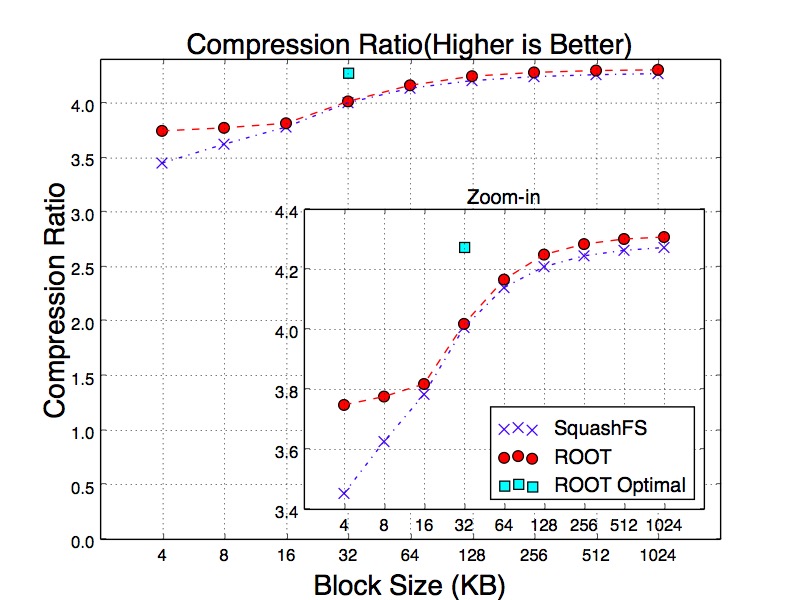}
\end{center}
\caption{Compression ratios of SquashFS and ROOT with various block sizes}
\label{fig:sqrshratio}
\end{figure}

\subsection{Results}

\begin{figure}[ht!]
\centering
\subfloat[Read size of all events]{\includegraphics[height=2in, width=2in]{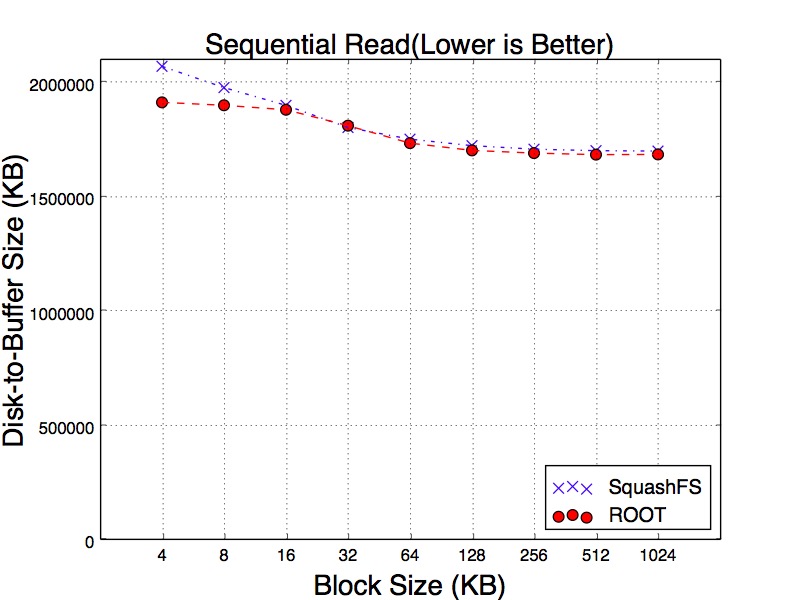}\label{fig:sqrshallsize}}
\subfloat[Read size of 10\% events]{\includegraphics[height=2in, width=2in]{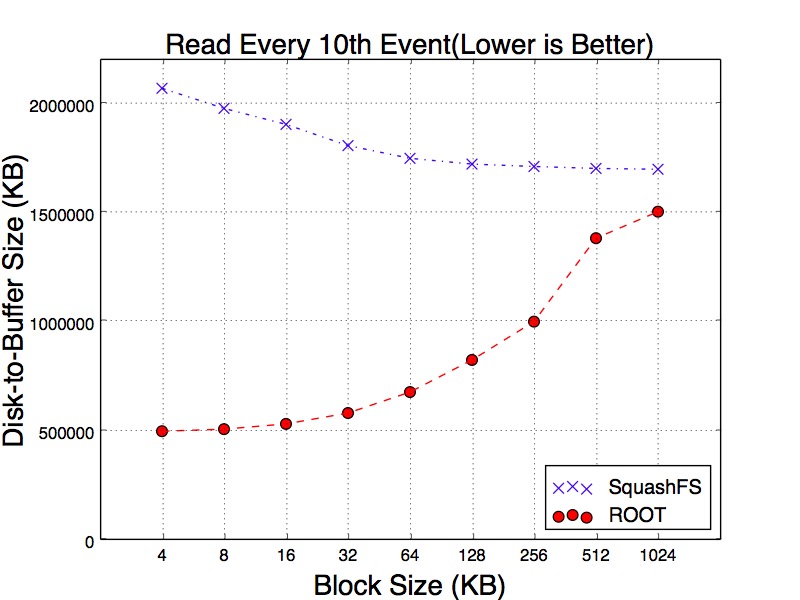}\label{fig:sqrsh10thsize}}
\subfloat[Read size of 1\% events]{\includegraphics[height=2in, width=2in]{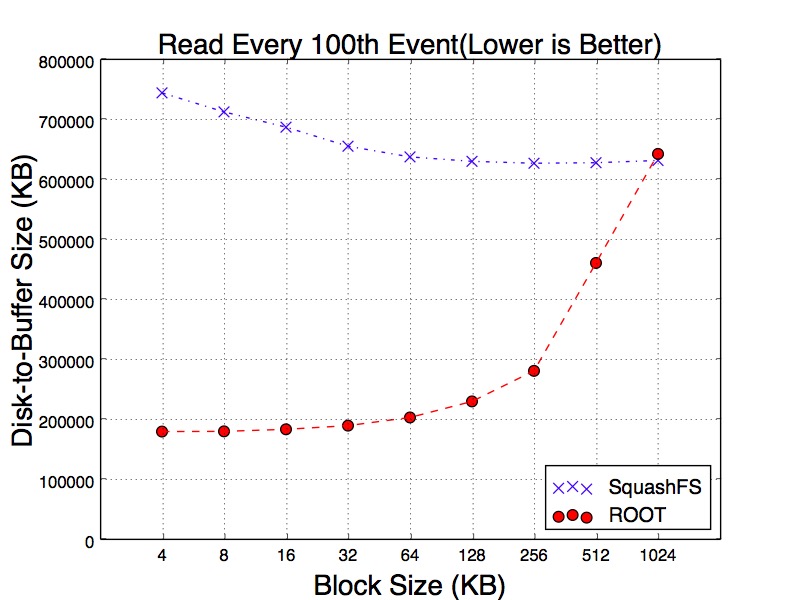}\label{fig:sqrsh100thsize}}\\[-0.5ex]
\subfloat[Read time of all events]{\includegraphics[height=2in, width=2in]{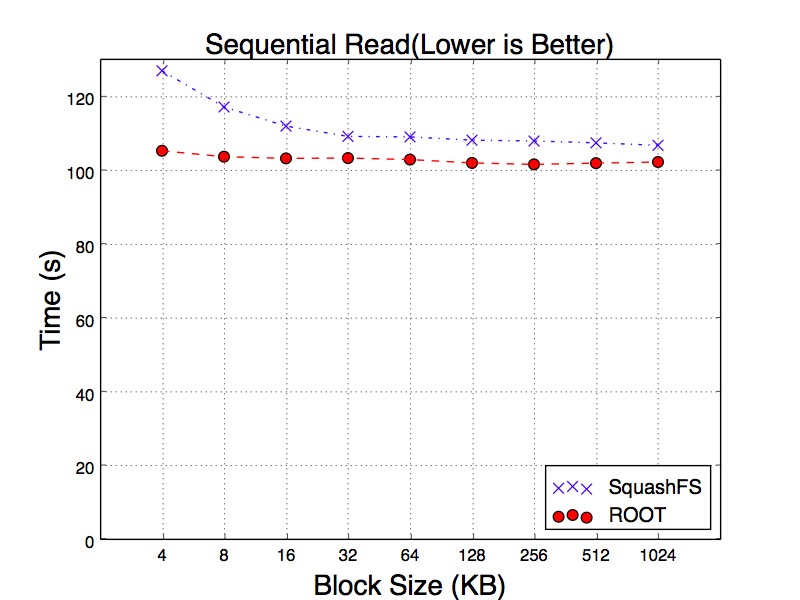}\label{fig:sqrshalltime}}
\subfloat[Read time of 10\% events]{\includegraphics[height=2in, width=2in]{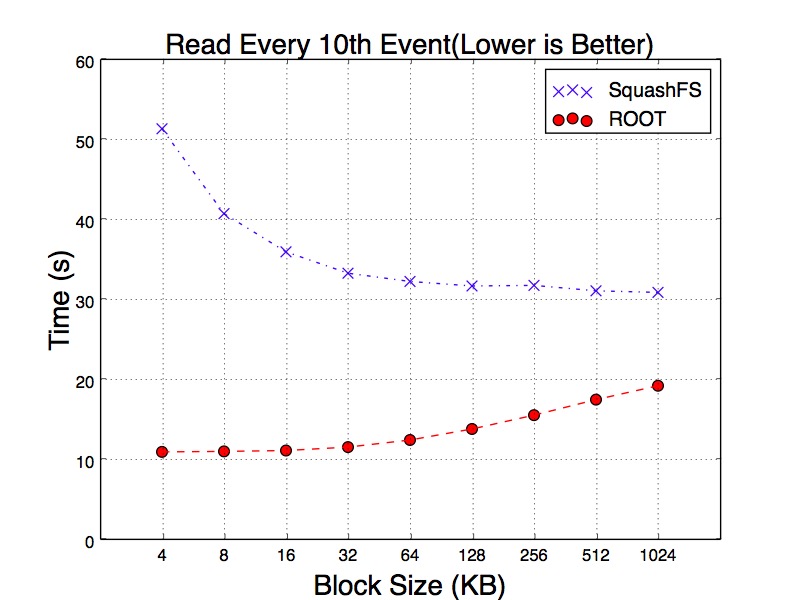}\label{fig:sqrsh10thtime}}
\subfloat[Read time of 1\% events]{\includegraphics[height=2in, width=2in]{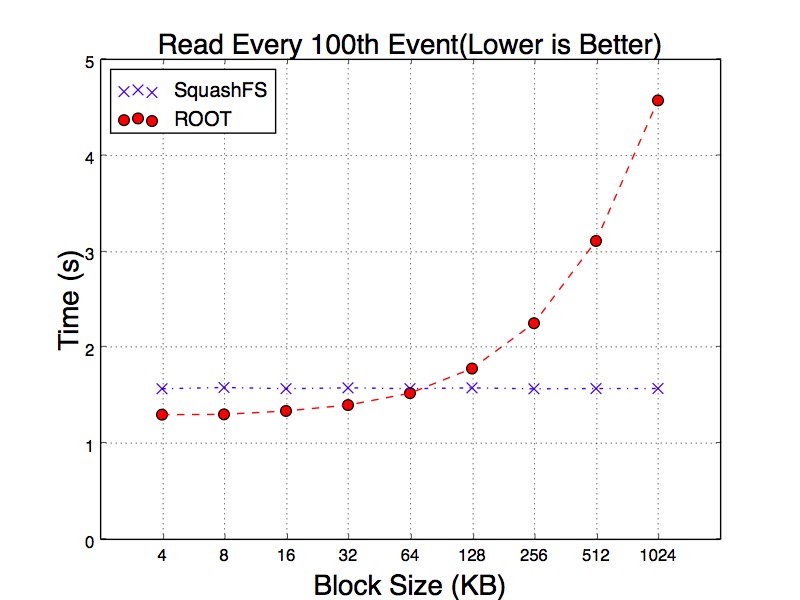}\label{fig:sqrsh100thtime}}
\caption{Disk-to-buffer sizes(cold cache) and read time(hot cache) of three workloads}
\label{fig:sqrshread}
\end{figure}

Figure \ref{fig:sqrshratio} illustrates compression ratios of SquashFS with different block sizes ranging between 4 KB to 1 MB. As the block size increases, compression ratio is improved. With the block sizes between 4KB and 16KB, ROOT has better compression ratios than SquashFS. When the block size goes beyond 16KB, their compression ratios become very close to each other. The result proves that ROOT organizes the data better than SqushFS. In order to get into more depth of the variance, we test three different reading workloads. The first workload is sequentially reading all events. The second is reading every 10th events (10\% of total events). The third workload reads every 100th events (1\% of total events). The purpose of this experiment is to simulate the sparse reads. We believe sparsely scanning the events can mitigate the cache effects. The results are shown in Figure \ref{fig:sqrshallsize}, \ref{fig:sqrsh10thsize} and \ref{fig:sqrsh100thsize}. Particularly, as can be seen in Figures \ref{fig:sqrsh10thsize} and \ref{fig:sqrsh100thsize}, SqushFS needs to fetch more data from disk in order to read the events since it does not know the data layout. One example could be an event is located across the boundary of two SquashFS blocks and reading this event requires to load two blocks from disk to the buffer.

Figures \ref{fig:sqrshalltime}, \ref{fig:sqrsh10thtime} and \ref{fig:sqrsh100thtime} show the time spent on running the same reading workloads. All tests have been running with hot cache. Figure \ref{fig:sqrsh10thtime} illustrates that SqushFS and ROOT go to the opposite trends as the block size gets larger. For SquashFS, the larger block can reduce the chance of loading multiple blocks for a single event. As a result, it takes less time to read the events. ROOT only reads the minimum number of baskets. However, as each of the baskets becomes larger, it has to spend more time on decompressing the data.

Figure \ref{fig:sqrsh100thtime} illustrates a more interesting result. Since SquashFS decompresses data in kernel space, with hot cache all requested events appear to be uncompressed in the memory. So increasing block size does not have any impact on reading time. In contrast, ROOT decompresses the data in user space. As a result, the reading time still increasing with larger blocks.


\section{Conclusions and Future Work}

Any compression scheme has both advantages and disadvantages. We believe our results show that the existing ROOT functionality provides reasonable performance (in CPU cost and storage saved) across a wide range of use cases. Our work in RAC, externally compressed data, and alternate compression algorithms show this ``middle ground" can be improved to meet the needs of specific use cases. Users need to choose the compressor for their particular applications.

As improved compression techniques yield significant savings for companies with large data stores, there are new and improved algorithms to evaluate each year. We plan to work with the ROOT team to bring best-of-breed algorithms into the software stack. Further, as hardware compression becomes available, we plan to investigate offloading this task from the CPU.

The RAC and external compression techniques allow us to compare performance against potential improvements. We believe the external compression results using SquashFS demonstrate ROOT IO compression is a competitive approach to minimizing total file size. The RAC work demonstrates some speedup in random access: we believe further improvements need to be done to better tradeoff improved random access and the cost of compression markers.



\section*{Acknowledgments}

This work was supported by the National Science Foundation under Grant ACI-1450323. This research was done using resources provided by the Holland Computing Center of the University of Nebraska.


\section*{References}
\begin{thebibliography}{9}
\bibitem{Brun199781} Brun R and Rademakers F 1997 \textit{Nucl. Instr. Meth. Phys. Res.} \textbf{389} 81
\bibitem{Deutsch:1996:ZCD:RFC1950} Deutsch P and Gailly J L 1996 \textit{ZLIB Compressed Data Format Specification Version 3.3} 
\bibitem{1055714} Ziv J and Lempel A 1977 \textit{IEEE Trans. Inf. Theory} \textbf{23} 337
\bibitem{4051119} Huffman D A 1952 \textit{Proc. IEEE} \textbf{40} 1098
\bibitem{lzma} URL \href{http://www.7-zip.org/sdk.html}{http://www.7-zip.org/sdk.html}
\bibitem{lz4} URL \href{http://lz4.github.io/lz4/}{http://lz4.github.io/lz4/}
\bibitem{cms} URL \href{http://cms.web.cern.ch}{http://cms.web.cern.ch}
\bibitem{188994} Agarwal R, Khandelwal A, Stoica I 2015 \textit{12th USENIX Symposium on Networked Systems Design and Implementation (NSDI 15)} 337
\bibitem{squashfs} URL \href{http://squashfs.sourceforge.net}{http://squashfs.sourceforge.net}

\end{thebibliography}
\end{document}